 \documentclass{aa} 
 \usepackage{amsmath}
\usepackage{txfonts}
\usepackage{graphicx}
\usepackage{natbib}
\usepackage{amsfonts}
\usepackage{amssymb}
\usepackage{hyperref}
\usepackage{xcolor}

\begin{document}

   \title{Magnetic activity cycles in solar-like stars: \\ The cross-correlation technique of p-mode frequency shifts}
   
  \author{C. R\'egulo\inst{1,2}
         \and R. A. Garc\'ia\inst{3}
          \and J. Ballot\inst{4,5}
          }

   \institute{Instituto de Astrof\'{\i}sica de Canarias, 38205, La Laguna, Tenerife, Spain
             \email{crr@iac.es}
              \and Universidad de La Laguna, Dpto de Astrof\'{\i}sica, 38206, La Laguna, Tenerife, Spain
              \and Laboratoire AIM Paris-Saclay, CEA/DRF-CNRS-Universit\'e Paris Diderot, IRFU/SAp, Centre de Saclay, 91191 Gif-sur-Yvette cedex, France
              \and CNRS, Institut de Recherche en Astrophysique et Plan\'etologie, 14 avenue Edouard Belin, 31400 Toulouse, France
              \and Universit\'e de Toulouse, UPS-OMP, IRAP, 31400 Toulouse, France}

   \date{Received ; accepted }
   
     \abstract
     {}
     {We aim studying the use of cross-correlation techniques to infer the frequency shifts induced by changing magnetic fields 
     in the p-mode frequencies and provide precise estimation of the error bars.}
     {This technique and the calculation of the associated errors is first tested and validated on the Sun where the p-mode
     magnetic behaviour is very well known. These validation tests are performed on 6000-day time series of Sun-as-a-star observations delivered by the SoHO spacecraft. Errors of the frequency shifts are quantified through Monte Carlo simulations. 
     The same methodology is then applied to three solar-like oscillating stars: HD~49933, observed by CoRoT, as well as
     KIC~3733735 and KIC~7940546 observed by \emph{Kepler}.}
     {We first demonstrate the reliability of the error bars computed with the Monte Carlo simulations using the Sun. From the 
     three analyzed stars we confirm the presence of a magnetic activity cycle with this methodology in HD~49933 and we 
     unveil seismic signature of on going magnetic variations in KIC~3733735. Finally, the third star, KIC~7940546, seems to be in a quiet regime.}
     {}
 
  \maketitle
  
  \section{Introduction}

The ultra-high precision photometric time series obtained with satellites such as CoRoT \citep[Convection Rotation and planetary 
Transits,][]{2006cosp...36.3749B}
and \emph{Kepler} \citep{2010Sci...327..977B} has opened a new window in the study of magnetic activity cycles in solar-like pulsating 
stars  \citep[e.g.][]{2013A&A...550A..32M,2014A&A...562A.124M,2014JSWSC...4A..15M}.  The active regions crossing the visible surface 
of a star produce a modulation in the observed light curve that provides information about the rotation rate and the stellar magnetic 
activity \citep[e.g.][]{2014A&A...572A..34G}. Moreover, it is even possible to detect magnetic frequency
shifts of the acoustic modes along the activity cycle \citep{2010Sci...329.1032G} as it is commonly done for 
the Sun \citep[e.g.][]{1985Natur.318..449W,1992A&A...255..363A,2001MNRAS.322...22C}.

To study the temporal evolution of the p-mode frequency shifts, it is necessary to extract reliable  p-mode frequencies of 
individual modes during short periods of time, typically 30 to 60 days. This could be a big challenge in asteroseismology, 
especially when changes of about a few $\mu$Hz are expected and the signal-to-noise ratio (SNR) could be small. An alternative 
way consists in computing average frequency shifts by using 
cross-correlation techniques. This method was first successfully used in the Sun with single site observations \citep[e.g.]
[]{1989A&A...224..253P}. With the arrival of long time series of hundreds of solar-like stars from space missions, the 
cross-correlation method may help us again in the study of magnetic activity cycles. Thanks to this technique, obtaining 
frequencies of individual p modes in short time series -- which requires high SNR -- is not needed. Instead, 
the cross-correlation of the full (or part) p-mode power excess (the frequency range in which the power of the 
modes is found) of each subseries and a reference spectrum can be computed providing the average
shift of the acoustic modes during the analyzed time span.

A problem in the use of the cross-correlation methods could be the way in which the individual uncertainties are computed.
In previous works in which this technique has been used, these uncertainties were obtained as the statistical error of
the fit of a Gaussian function to the cross-correlation
profile  \citep[e.g.][]{1989A&A...224..253P,2010Sci...329.1032G,2013A&A...550A..32M}. This fit is the way to determine the average 
frequency shift between two different periods of time. Such errors underestimate total errors since systematic and model 
errors are neglected. To better quantify the errors, we propose in the present work to use Monte Carlo simulation. We analyze 
the limit of the methodology using 6000 days of solar data obtained with the blue channel of the solar photometer of VIRGO 
instrument \citep{1995SoPh..162..101F} onboard the SoHO spacecraft \citep{DomFle1995}.

The method is applied to three solar-like stars: one from CoRoT, HD~49933, and two from \emph{Kepler},  
KIC~3733735 and KIC~7940546. It is well known that the first two stars have on-going magnetic 
cycles \citep{2010Sci...329.1032G,2014A&A...562A.124M} while the third star, KIC~7940546, does not show any significant 
variation in the light curve which could be a consequence of a low inclination angle or because the star is in a quiet magnetic state.

The layout of the paper is as follows. In section 2 we will explain the methodology used. In section 3 the method is validated using 
6000-day time series of Sun-as-a-star observation. In section 4 three solar-like stars, one from CoRoT and two from \emph{Kepler} 
satellite are analysed. The last section is devoted to discussion and conclusion. 

\section{Method}

We use the cross-correlation technique to measure  p-mode frequency shifts.
To use this method, we first divide the whole light curve of the analysed stars into shorter subseries that can overlap or not. 
The size of each chunk depends on the overall length of the observations, the quality of the data, and the length of the activity 
cycle we want to investigate. According to these restrictions, we have used series between 30 and 180 days for the three 
analysed solar-like stars. For the Sun we use longer subseries --up to a year-- because we have more 
than 16 years of observations and because the solar magnetic cycle is long (eleven years in average from maximum to maximum).  
The power spectrum of each subseries is computed.

Then, each one is cross-correlated with a reference spectrum defined as the average of all the individual spectra. 
The profile of the cross-correlation function has almost a Gaussian shape. Therefore, a Gaussian function is fitted to each 
cross-correlation function. The position of the maximum is considered as the average shift of the p-mode frequencies present 
in the range of the spectrum used in the analysis.

To obtain the error bars of the cross-correlation analysis we follow the following steps:

1 - At high frequencies,  well above the p-mode region where the spectrum is dominated by the photon noise, 
the standard deviation  of this noise is obtained for both spectra, the reference one and the target one. 

2 - The absolute difference between the standard deviations of the target spectrum and the reference one 
is added to the later in order to have both spectrum with the same noise level.

3  - The reference spectrum with the new level of white noise is shifted by the computed shift (if any).

4 - We produce one hundred simulated power spectrum by multiplying the reference shifted spectrum by a random noise 
distribution following a  $\chi^{2}$ with 2 degrees of freedom.

5 - Each of these simulated spectra is cross-correlated with the original reference one to obtain a new frequency shift.

6 - The standard deviation of all the obtained displacements of the simulated spectra with respect to the real one is 
considered the error of the measure.

We have verified that the results are unbiased by comparing the mean extracted values to the expected one.
We have also verified that using an averaged spectrum as a reference spectrum is not introducing significant biases.
Moreover, we checked that a higher number of simulated realizations (e.g. 500 or 1000) does not change the final result. 
Therefore we have decided to use 100 realizations to reduce the computation time.

\section{Validation: application to the Sun}

\begin{figure}[!htb]
 
  \resizebox{\hsize}{!} {\includegraphics[angle=90]{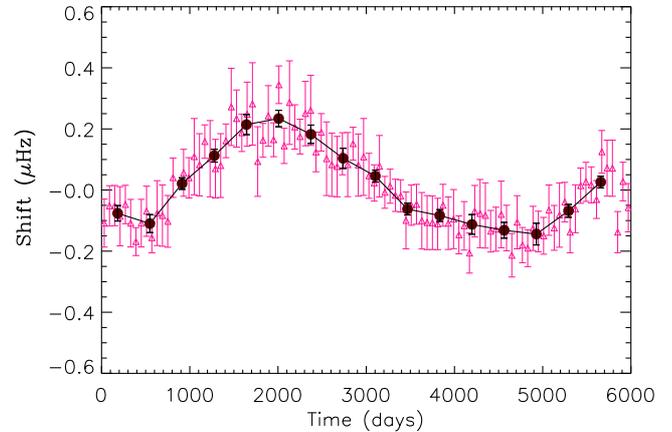}}
   \caption{Frequency shifts along the solar cycle using subseries of 60 (open triangles) and 365 days 
   (full circles). The error bars 
   have been obtained using  Monte Carlo simulations following the methodology described in this work. Starting on April $11^{\rm{th}}$ 1996.}
              \label{fig:shift_sun}
 \end{figure}

To study the behaviour and the limits of the cross-correlation analysis as well as the error estimation we use the Sun as 
a reference. Time series of 6000 days cadenced at 60 s --from April $11^{\rm{th}}$ 1996 to September $14^{\rm{th}}$
2012--  are used in this analysis almost covering one and half solar magnetic cycle. Solar observations were obtained by the blue 
channel of the Sun Photometers of the VIRGO instrument onboard the SoHO spacecraft. With these observations we study 
the behaviour of the method attending the range of p-mode frequencies used, the length of the subseries, the influence 
of a pre-smooth of the spectra, the SNR, and the width of the modes. For doing so we divided the time series into
independent subseries (without any overlap) of 30, 60, 90, 120, 180, 240, 300 and 365 days. An example of the resultant 
frequency shifts using 60 and 365 days subseries is shown in Fig.~\ref{fig:shift_sun}.

\begin{figure}[!htb]
 
  \resizebox{\hsize}{!} {\includegraphics[angle=90]{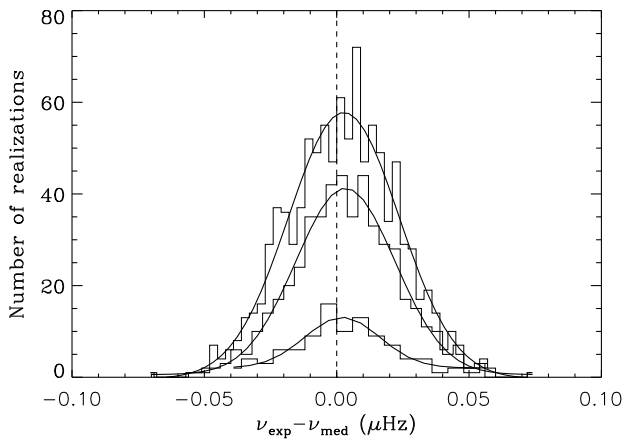}}
   \caption{Histogram with fitted Gaussian profiles, of the differences between the expected  and the measured shifts in the Monte Carlo 
   simulation for 100, 500 and 1000 realizations. Series of 365 days are used.
}
              \label{fig:histograms}
 \end{figure}

 As a first step, we checked that the computation of the errors are unbiased by calculating a histogram of the difference
 between the expected shift ($\nu_{exp}$) and the measured shifts ($\nu_{med}$) in the Monte Carlo simulations. This is done for series of 
 365 days and the results are shown in Fig.~\ref{fig:histograms} for 100, 500 and 1000 realizations. As we expect from an 
 unbiased computation, 
 the histograms are well fitted by a Gaussian profile centred in zero. In our case, the small deviation from 
 zero that the Gaussian profiles shown is 0.003 $\mu$Hz, inside the dispersion of 0.005 $\mu$Hz obtained from the errors associated to 
 the frequency shifts calculated from the series of 365 days.

To verify that using an averaged spectrum as a reference spectrum is not introducing significant biases, 
we create a synthetic limit spectrum miming the solar spectra observed by VIRGO within the range 2000--4000~$\mu$Hz. 
It includes a background and modes of degree $l=0$ to 3. We consider two spectral resolutions corresponding to 60-day- and 365-day-long 
observations. For each resolution, we generate 100 synthetic observations by multiplying the limit spectrum by a random noise 
distribution following a  $\chi^{2}$ with 2 degrees of freedom. Then, we consider sequences of 100, 50, 20, 10, 5 or even 2 independent 
observations. Each sequence is analysed with the previously described technique (measurement of the shift -- expected to be zero -- and 
of the error bar). We compare the results obtained by using directly the limit spectrum as a reference instead of an average spectrum.
We don't see any significant difference. Only when the number of spectra in the sequence is small (2 or 5), we see two marginal effects: 
a slight bias of the determined shift towards zero and a slight increase of the estimated error bars. However, these effects are weak 
compared to the errors themselves.

\begin{figure}[!htb]

  \resizebox{\hsize}{!} {\includegraphics[angle=90]{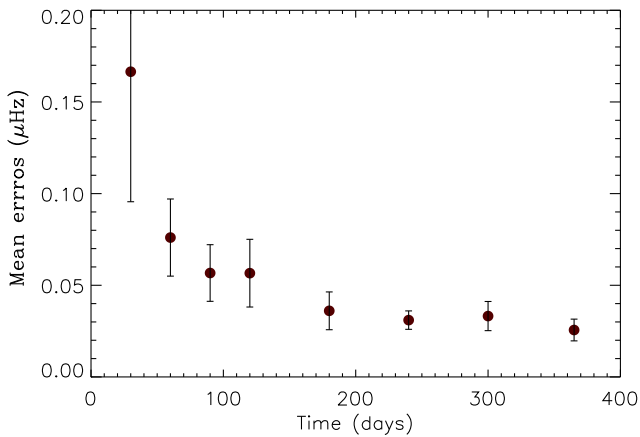}}
   \caption{Mean errors and associated dispersions ($\sigma_{\mathrm{error}}$) of the frequency shifts as a function of the length of the subseries.}
              \label{fig:length}
    \end{figure}

After that, we start analyzing the behaviour of the computed errors depending on the length of the subseries following the method 
explained in the previous section. In Fig.~\ref{fig:length} it is shown the average errors and their 
dispersions ($\sigma_{\mathrm{error}}$) as a function of the length of the subseries. The frequency shifts
were computed using a range of $\pm$ 600 $\mu$Hz centered at 3000 $\mu$Hz and without smoothing the spectra. 
The errors decrease with the increase of the length of subseries $T$. It varies as $T^{-0.65 \pm 0.14}$ which 
is almost compatible with a squared-root decay.

\begin{figure}[!htb]
 
  \resizebox{\hsize}{!}  {\includegraphics[angle=90]{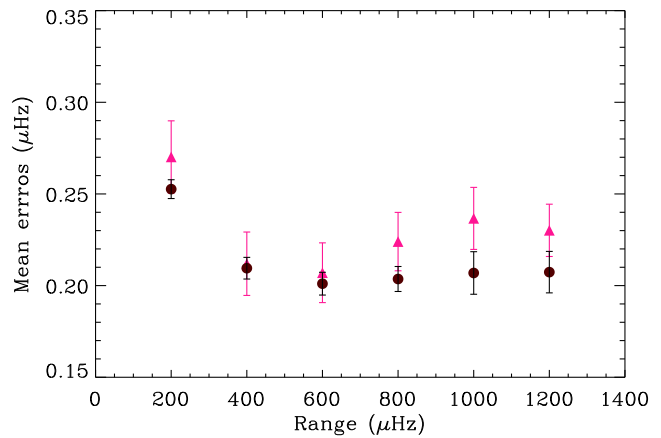}}
   \caption{Average errors and associated dispersions of the frequency shifts using series of 365 days and different 
   ranges of p modes. Triangles correspond to time series with a SNR=1.1, while circles correspond to the nominal 
   VIRGO/SPM observations (SNR=3.3). In this later case the average errors were multiplied by 7 to adjust the scale.}
   \label{fig:range}
    \end{figure}

The fact of using this range of $\pm$ 600 $\mu$Hz is because it is a good compromise between the size and dispersion of 
the obtained errors, as it can be seen in Fig.~\ref{fig:range}. Indeed, different ranges of p modes centred at 
3000 $\mu$Hz: $\pm$ 200, 400, 600, 800, 1000, and 1200 $\mu$Hz have been compared. Two sets of results were plotted. 
Triangles correspond to the results obtained when white noise is added to the time series to reproduce noisy data 
with a SNR three times lower than the original (SNR=1.1). The SNR is estimated as the mean value of the ratio between the mode 
amplitude in the power spectra and the fitted background in the analyzed
range. Circles correspond to the original signal (SNR = 3.3). 
Although the results are quite similar, smaller errors are obtained when a range of $\pm$ 600 $\mu$Hz around the central 
frequency of 3000 $\mu$Hz is used. 

\begin{figure}[!htb]
 
  \resizebox{\hsize}{!} {\includegraphics[angle=90]{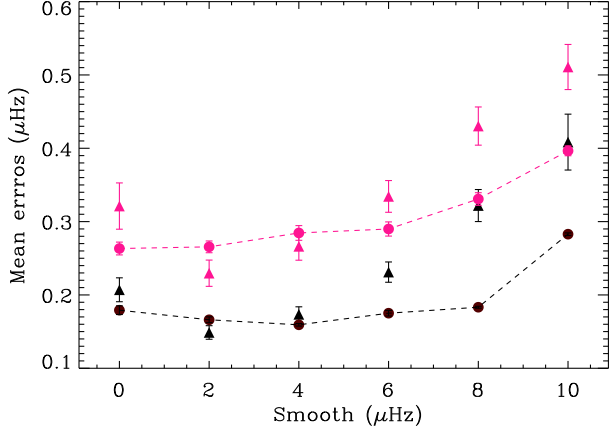}}
   \caption{Average errors and dispersions of the frequency shifts as a function of the smooth factor of the power 
   spectrum for series of 365 days (black) and 180 days (pink)  using a range of $\pm$600 $\mu$Hz centred at 3000 $\mu$Hz. 
   Same symbol code as 
   in Fig.\ref{fig:range} for circles and triangles. The results from the original series (without noise added) have been 
   connected by dashed lines.}
   \label{fig:smooth}
    \end{figure}

The next step is to analyze the behaviour of the errors when the spectra are smoothed before computing the cross 
correlation. In Fig.~\ref{fig:smooth} we plot the mean errors and their dispersions for series of 365 and 180 days using
$\pm$ 600 $\mu$Hz centred in 3000 $\mu$Hz, but with different levels of smooth. The number of points used for the 
smooth are: 65, 125, 191, 251, and 315, corresponding approximately with length of: 2, 4, 6, 8, and 10 $\mu$Hz. 
The smaller errors and dispersions are obtained when smoothing factors of 65 or 125 points are applied, corresponding 
to smoothing windows of 2 or 4 $\mu$Hz, which depends on the amount of noise of the time series. Once again, triangles
correspond to noisy data (SNR = 1.1) and circles to the original time series (SNR = 3.3). The series of 365 days are plotted
in black and the series of 180 days in pink. It is 
interesting to emphasize that larger smooth degradate the results, probably because it smears out the spectra
too much, thus attenuates the effect of searched shifts.

To study the behaviour of the cross-correlation technique with different levels of noise and different p-mode widths, 
the mean spectrum of 365 days has been fitted to obtain the width of the individual p modes and the SNR of the spectrum. 
The fit was done in a global way  using a Maximum Likelihood Estimatior \citep[e.g.][]{2008A&A...488..705A}, where the 
same width is used for each group of one l = 0, 1, and 2 modes. The so-called ``noise background'' \citep[e.g.][]{2011ApJ...741..119M} 
was modelled with three components: a flat white noise component dominating the high-frequency part of the spectrum, a power law 
at low frequency, and one Harvey component \citep{harvey85} to take into account the granulation contribution. This noisy background
is fitted first and then fixed when the p modes are fitted as it is usually done in asteroseismology
\citep[e.g.][]{2012A&A...543A..54A,2013A&A...549A..12M}. The errors in the fitted parameters were obtained from the Hessian matrix.

\begin{figure}[!htb]
 
  \resizebox{\hsize}{!} {\includegraphics[angle=90]{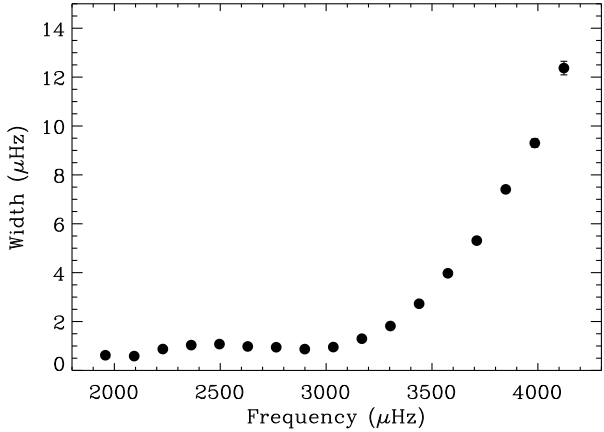}}
   \caption{Plot of the solar p-mode widths as a function of frequency.}
   \label{fig:width}
\end{figure}

The resultant mode  widths as a function of frequency is shown in Figs.~\ref{fig:width}.
The spectrum has p modes with widths less than 1 $\mu$Hz below 3000 $\mu$Hz  and wider than 2 $\mu$Hz above 3300 $\mu$Hz.

We have computed the cross-correlation for the spectra of 365 days in two different ranges, from 2100 to 2900 $\mu$Hz
and from 3300 to 4100 $\mu$Hz without smooth and with a smooth of 2 $\mu$Hz (65 points). In these two ranges the mean SNR is 2.0 for 
the first range and 2.9 for the second. 

\begin{figure}[!htb]
 
  \resizebox{\hsize}{!} {\includegraphics[angle=90]{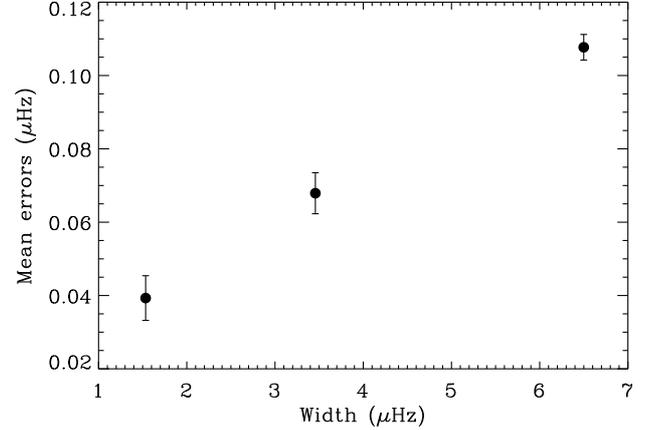}}
  \caption{Errors in the frequency shifts and associated dispersions as a function of the p-mode widths for series
  of 365 days with no smooth and centered at 3200, 3500 and 3800 $\mu$Hz in a range of $\pm$ 300 $\mu$Hz.}
              \label{fig:ranges}
    \end{figure}

As it is well known \citep[e.g.][]{2004A&A...413.1135S}, the frequency shift is higher for the Sun at high frequencies. 
At the same time, the SNR is better at high frequencies, however, when the mean values of the errors are obtained 
for both ranges, see Table~\ref{table:ranges}, the errors 
are $\approx$ 60\% higher at high frequencies. This can only be attributed to the higher width of the modes at high 
frequencies because the height of the modes are similar in both regions. In the range where the width of the 
p modes increase, we have analyzed the errors obtained in the computation of the frequency shifts for series of 365 
days with no smooth and centered at 3200, 3500 and 3800 $\mu$Hz in a range of $\pm$ 300 $\mu$Hz. The result is shown
in Fig.~\ref{fig:ranges}. The errors increase with the mode 
width $\Gamma$ as  $\Gamma^{0.7\pm0.1}$. We expect that the variation goes as $\Gamma^{0.5}$ following similar 
argument as \citet{1992ApJ...387..712L}. Nevertheless, we have to be cautious for two reasons. First, 
we must keep in mind we only get three points, which is a small number to derive a fine quantitative law. Second, 
the squared-root dependency is only a lower limit of the errors.

\begin{table}
\caption{Obtained frequency shifts for two different computed regions and smooth factors (no smooth, s0, and a smooth 
window of 65 points, s65). The printed shift is the peak to peak value obtained in the cross-correlation
analysis and the error is the mean value of the obtained errors. All quantities are in $\mu$Hz.}
\label{table:ranges}
\centering 
\begin{tabular}{c c c c  c} 
\hline\hline
Range   & shift (s0) & error (s0) & shift (s65) & error (s65)\\
\hline
2100 - 2900      &  0.209     & 0.032     & 0.235       & 0.035 \\
3300 - 4100     &   0.540     & 0.054     & 0.553       & 0.053 \\

\hline
\end{tabular}
\end{table}

\begin{figure}[!htb]
 
  \resizebox{\hsize}{!} {\includegraphics[angle=90]{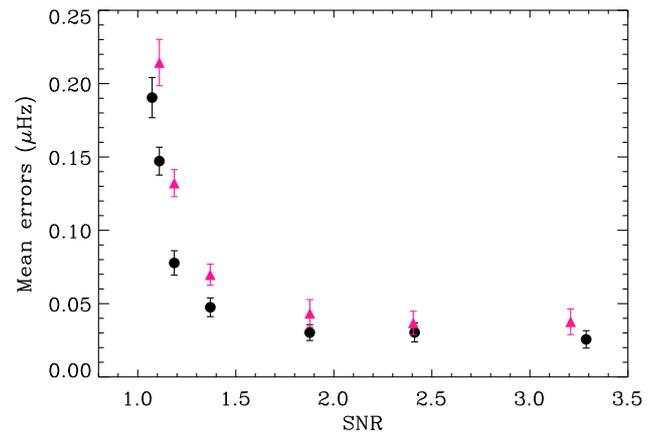}}
   \caption{Average frequency shift errors against the SNR. The range used for the calculation is from 2400 to 
   3600 $\mu$Hz and no smooth has been applied. Circles correspond to series of 365 days and triangles to series of 180 days.
 }
              \label{fig:SNR}
    \end{figure}

To study the influence of the SNR on the cross-correlation analysis we have added different levels of normal 
white noise to the origianl VIRGO/SPM time series of 6000 days. The cross-correlation has been computed between 
2400 and 3600 $\mu$Hz and no smooth has been applied. Results are shown in Fig.~\ref{fig:SNR}. The errors drop 
drastically with the increase of 
SNR until reaching SNR=2. For higher SNR, the reduction is very small.

\section{Application to solar-like stars}

\subsection{The CoRoT target HD~49933}

HD~49933 is a F5 main-sequence star observed twice by CoRoT, with a first run of 
60 days in 2007 and a second of 137 days in 2008. The light curves were obtained
with a cadence of 32 seconds. It is a star hotter than the Sun, with an effective temperature around 6700 K, an estimated 
mass of 1.3 solar mass, and an estimated radius of 1.34 solar radius. This star 
has been widely analysed \citep[e.g.][]{2008A&A...488..705A,2009A&A...507L..13B,2010A&A...510A.106K}, and it was also 
the first star, apart from the Sun, in which asteroseismic measurements of global changes in the frequencies of their 
acoustic modes were reported \citep{2010Sci...329.1032G,2011A&A...530A.127S}, confirming the development of an on-going 
stellar magnetic activity cycle.

In this work we use the series of 137 days long to analyze the behaviour of the errors of the cross-correlation signal 
using Monte Carlo simulations. In particular, we have used series of 30 and 60 days shifted every 15 days. The reference 
spectrum is the average spectrum obtained from independent series, i.e., 4 spectra are used in the case of 30-day long
subseries, while only two subseries are averaged when the length of the subseries is 60 days.  The power spectrum of 
each subseries is centered at $\nu_{\rm{max}}$=1850 $\mu$Hz (the frequency of maximum power), and we study different 
frequency ranges, as well as smooth factors. The number of points used to smoothing the spectrum are 11, 21, and 41, 
corresponding to smoothing windows of 2, 4 and 8 $\mu$Hz in the case of the subseries of 60 days. It is important to
note that he smaller errors are obtained with the 11 points smoothing factor.

A summary of the results obtained for the different ranges using subseries of 60-day long with an 11-point smoothing 
factor is shown in Table~\ref{tab:hd49933}. Using the range of $\pm$400 $\mu$Hz around $\nu_{\rm{max}}$ we obtain the 
smallest errors in the computed frequency shifts.

 \begin{table}
\caption{Average errors obtained in the analysis of the 60-day subseries covering different frequency ranges for HD~49933. 
The cross-correlated spectra have been smoothed using 11 points. All units are in $\mu$Hz.}
 \label{tab:hd49933}
\centering 
\begin{tabular}{c c c} 
\hline\hline
    Range &Mean error   &    $\sigma_{error}$\\
    \hline
   $\pm$ 200      &   0.270  &   0.002\\
   $\pm$ 300      &   0.258  &   0.019\\   
   $\pm$ 400      &   0.215  &   0.011 \\     
   $\pm$ 500      &   0.222  &   0.006 \\      
   $\pm$ 600      &   0.233  &   0.037\\
   \hline
\end{tabular}
\end{table}

\begin{figure}

  \resizebox{\hsize}{!} {\includegraphics[angle=90]{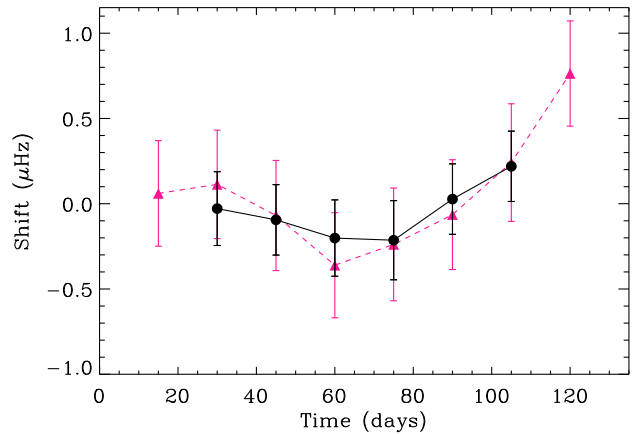}}
   \caption{Frequency shifts obtained for HD~49933 computed with subseries of 30 days (triangles) and 60 days 
   (circles) shifted every 15 days. The region used is centred at $\nu_{\rm{max}}$=1850 $\mu$Hz with a range 
   of $\pm$400 $\mu$Hz and a smoothing factor of 11 points.}
              \label{fig:freqhd}
    \end{figure}

The temporal evolution of the frequency shifts of HD~49933 is shown in Fig.~\ref{fig:freqhd}  for series of 30 and
60 days centred in $\nu_{\rm{max}}$=1850 $\mu$Hz with a range of $\pm$400 $\mu$Hz and a smoothing factor of 11 points. 
These results are in well agreement with the obtained by \citet{2010Sci...329.1032G}.

In this star, the width of the p modes is  between 3 and 11 $\mu$Hz \citep{2009A&A...507L..13B}. In the range 
of $\pm$ 400 $\mu$Hz around $\nu_{\rm{max}}$, the average p-mode width is 6.3 $\mu$Hz, and the SNR for the 
60-day average spectrum is 2.1, see Table~\ref{tab:SNR}.

As we expected from the width of the acoustic modes and the level of noise, the computed errors are quite big. 
The use of 60 days subseries produces smaller errors. However, when the longer subseries are used the resultant 
maximum frequency shift is reduced, due to the short period of the magnetic activity cycle of the star and the short 
amount of data available (137 days). Indeed the maximum frequency shifts are 1.12 $\pm$ 0.32 $\mu$Hz and 0.43 $\pm$ 0.21 $\mu$Hz
respectively for the calculation using 30 and 60-day subseries.

 \begin{table}
\caption{SNR and mean width in the analysed range, for all the analysed stars}
 \label{tab:SNR}
\centering 
\begin{tabular}{c c c} 
\hline\hline
    Star       & SNR     & Mean width ($\mu$Hz)\\
    \hline
   Sun         &   3.3     &  1.6  \\
   HD 49933    &   2.1     &  6.3   \\   
   KIC 3733735  &  1.7     &  8.9  \\     
   KIC 7940546  &  3.0     &  3.5  \\      
   \hline
\end{tabular}
\end{table}

\subsection{The \emph{Kepler} targets KIC~3733735 and KIC~7940546}

KIC~3733735 is also a F5 main-sequence star, observed by \emph{Kepler} in short cadence
\citep[58.85 seconds,][]{2010ApJ...713L.160G,2011MNRAS.414L...6G} along 1114 days. A seismic study has been 
performed by \citet{2012A&A...543A..54A}, and its magnetic activity was analyzed by \citet{2014A&A...562A.124M},
where a cycle-like behaviour has been detected from the study of the temporal evolution of the light curve. 
KIC~3733735 has a $T_{\rm{eff}}$ around 6700 K, a radius of 1.66 solar radius, and a mass of 1.95 solar masses
\citep{2012ApJ...749..152M}. The acoustic modes of this star are wider than the modes of the previous analyzed
CoRoT star, from 4 to 14 $\mu$Hz  \citep{2014A&A...566A..20A}. Thus, the mean linewidth is 8.9 $\mu$Hz 
 in the range 1650 - 2250 $\mu$Hz.

\begin{figure}[!htb]
 
  \resizebox{\hsize}{!} {\includegraphics[angle=90]{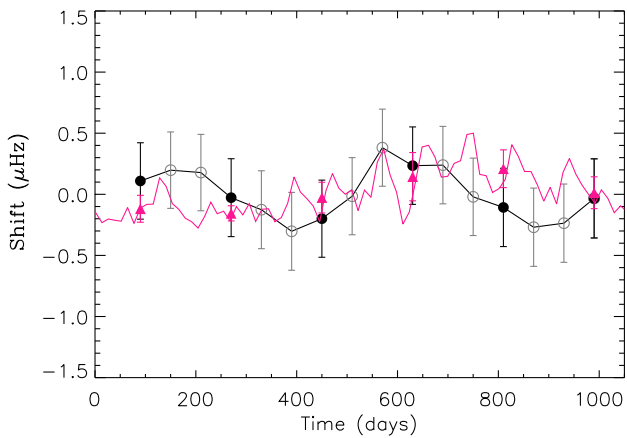}}
   \caption{Frequency shifts of KIC~3733735 calculated using 180-day subseries shifted every 60 days.The points obtained 
   from independent spectra are shown as full black circles. The power 
   spectrum is centred at a $\nu_{\rm{max}}$=1950 $\mu$Hz with a range of $\pm$ 300 $\mu$Hz and a smooth of 31 
   points. The mean spectra used as the reference for the cross-correlation are generated with independent spectra. 
   The pink curve is the $S_{ph}$ computed by \citet{2014A&A...562A.124M}. The pink triangles are the integration of the $S_{ph}$ 
   to the same 180d ranges than the frequency shifts.  }
              \label{fig:sk}
    \end{figure}

In Fig.~\ref{fig:sk} we show the computed frequency shifts using subseries of 180 days shifted every 60 days. 
Each spectrum is centred in $\nu_{\rm{max}}$=1950 $\mu$Hz, with a range of $\pm$ 300 $\mu$Hz and smoothed by 
31 points, corresponding to a smoothing window of 2 $\mu$Hz (black curve). The reference spectrum is the mean of the independent 
spectra. This combination of parameters is the one producing the smallest errors. The SNR of the mean 180-day
spectrum in the selected range of p modes is 1.7. 

As the level of the noise and the width of the modes is something that we can not change in the observations, 
we can only check the effects of changing the length of the subseries, the p-mode range, and the smoothing factor. 
Similar to the solar case, the length of the series is the dominant factor producing a bigger change in the computed errors. 

For this star, and using the optimal configuration of a range of $\pm$ 300 $\mu$Hz and a smooth of 31 points, 
series of 30, 60, 90, and 180 days have been computed and the obtained mean errors from the cross-correlation with 
their associated dispersions are shown in Table~\ref{tab:sk}. The errors decrease 
with the length of the subseries similar to the solar case .

 In Fig.~\ref{fig:sk}, superimposed to the computed frequency shift, is the time evolution of the photospheric activity 
proxy \citep[$S_{ph}$,][]{2014JSWSC...4A..15M} obtained from the long cadence \emph{Kepler} data shown in the
bottom panel of Fig. ~1 by \citet{2014A&A...562A.124M}. We have integrated those values to the same 180 days than 
in the frequency shifts (pink triangles) for the independent series, that are shown as black full circles 
in the frequency shift plot. The error bars of these points correspond to the dispersion of $S_{ph}$  
in each 180-d segment. Comparing both proxies of the activity we see they are compatible inside the error bars but 
we observe a lag of about 120 days between the two. Several authors \citep[e.g.][]{2009ApJ...695.1567J,2009A&A...504L...1S,2015A&A...578A.137S} already 
discussed the possibility that there could be a time lag between the manifestation of the magnetism in the photosphere 
and the perturbation in the frequency of the acoustic modes.

 \begin{table}
\caption{Mean errors of the frequency shifts of KIC~3733735 and associated errors for series of 30, 60, 90,and 180 days.}
 \label{tab:sk}
\centering 
\begin{tabular}{c c c} 
\hline\hline
Days   &      Mean error ($\mu$Hz)  &   $\sigma_{error}$($\mu$Hz)\\
\hline    
30     &      1.099        & 0.063  \\    
60     &      0.669        & 0.029   \\   
90     &      0.476        & 0.023 \\   
180    &      0.278        & 0.017\\
   \hline
\end{tabular}
\end{table}

KIC~7940546 is the second \emph{Kepler} star that we have analyzed. It is also a F5 main-sequence star with 
a $T_{\rm{eff}}$ around 6350 K, a radius of 1.95 solar radius, and a mass of 1.41 solar mass. Its surface magnetic 
activity was also studied by \citet{2014A&A...562A.124M}, and it was one of the less active star in their sample. 

KIC~7940546 was observed in short cadence along 928 days.  Spectra from series of 30, 60, 90, and 180 days were
produced to obtain the frequency shifts in a range of $\pm$400 $\mu$Hz around $\nu_{\rm{max}}$=1080 $\mu$Hz. 
In Table~\ref{tab:kick2} the average errors in the frequency shifts and their dispersions obtained from the 
different subseries' lengths are shown. As we can notice, the mean errors are very small compared with those 
obtained for KIC~3733735 (see Table~\ref{tab:sk}) and they are comparable with the solar ones, (see Fig.~\ref{fig:length}).

 \begin{table}
\caption{Mean errors for KIC7940546 frequency-shift with their 
   dispersion ($\sigma$) for series of 30, 60, 90,and 180 days.
   The used range in the spectra is  $\pm$400 $\mu$Hz $\nu_{\rm{max}}$. No smooth has 
   been aplied to the spectra.}
 \label{tab:kick2}
\centering 
\begin{tabular}{c c c} 
\hline\hline

Days   &      Mean error ($\mu$Hz)  &   $\sigma_{error}$($\mu$Hz)\\
\hline    
30     &      0.195       & 0.027  \\    
60     &      0.157        & 0.023   \\   
90     &      0.112        & 0.011 \\   
180    &      0.074        & 0.004\\
   \hline
\end{tabular}
\end{table}

The seismic study of this star has not yet been performed. Therefore,  we do a similar global analysis as the
one described for the Sun. We found a mean width of the p modes of 3.5 $\mu$Hz and a very good SNR of 3.0. From 
these values, we effectively expect small errors.

\begin{figure}
 
  \resizebox{\hsize}{!} {\includegraphics[angle=90]{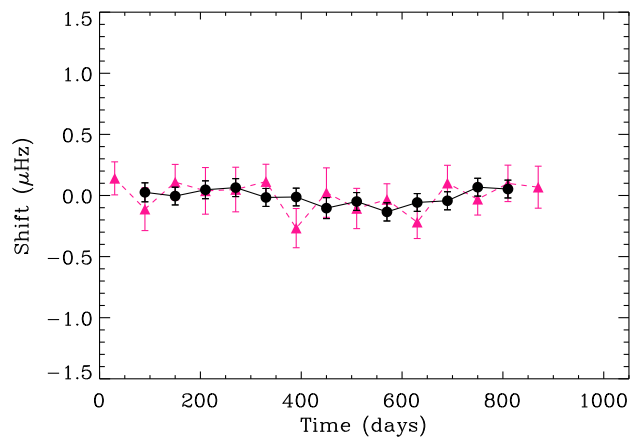}}
   \caption{Frequency shifts of KIC~7940546 calculated using independent 60-day subseries centred 
   in $\nu_{\rm{max}}$=1080 $\mu$Hz and with a range of $\pm$400 $\mu$Hz (triangles), and using 
   180-day subseries shifted 60 days (circles). The reference spectra is the average of the independent 
   subseries. In both cases the spectra has not been smoothed.}
              \label{fig:kick2}
    \end{figure}

In Fig.~\ref{fig:kick2}, we plot the obtained frequency shifts for the independent subseries of 60 days and
for the subseries of 180 days shifted every 60 days. As for the previous stars, the reference spectrum for the
calculation of the cross-correlation is the
mean spectrum obtained from independent subseries of data. As we expect from the magnetic analysis performed 
by \citet{2014A&A...562A.124M}
this star does not show any frequency shifts during all the observing time.

\section{Discussion and Conclusion}

The cross-correlation technique is a simple, fast, and non demanding method to implement in order to unveil 
the possible existence of on-going stellar magnetic activity cycles. This is particularly interesting when we
want to study hundreds to thousands stars with the present and future space instruments, because we can avoid 
the process of fitting the individual p modes of a huge number of stars, and this for every subseries. 

As the significance of the solution depends on the calculated errors, in the present paper, 
we derive a method to properly compute the errors of the frequency shifts using Monte Carlo simulations. We 
have first tested the methodology  using  the Sun and then with 3 well known solar-like stars observed with CoRoT and \emph{Kepler}.

We have found that the errors fall following a power law with the length of the series and with the SNR, while 
they increase with the width of the p modes.
 
Related to the length of the subseries used, they are limited by two factors: the length 
of the data and the length of the activity cycle. In several cases, a small level of smooth in the power spectrum
seems to help in reducing the final errors of the frequency shifts. 

In the case of the Sun, where we can use subseries as long as one year with  high SNR, we have managed to obtain 
errors as small as 
0.02 $\mu$Hz in a signal with a global change of 0.45 $\mu$Hz. That is, an error of $\sim$ 4\% of the signal.

In the case of HD~49933, the maximum length we managed to use was 60 days because the length of the observations 
was 137 days. For this star we obtain errors of 48\% of the global shift. These high errors are principally related
with the short lenghts of the used series.

In the case of KIC~3733735, the maximum length of the used series was 180 days because the change in the activity 
is no longer than 250 days from maximum to minimum. The obtained errors for this star were around 45\% of the signal. 
This big errors are due to the small SNR and the large widths of the p modes.

For comparison, we have also analysed a star, KIC~7940546, which is not showing any surface magnetic activity cycle 
during the observing time. Moreover, this star has not very broad p modes and has a very good SNR around $\nu_{\rm{max}}$. 
Thus, the obtained errors were small. The computed frequency shifts are zero inside 1 $\sigma$ of the errors that are as
small as we expected from the characteristic of 
the spectra.

Even with the obtained error levels, the difference in the results between the active and the quiet star are clear enough to justify the use of the cross-correlation technique as a fast way to analyse the mean 
frequency shifts of the huge amount of solar-like stars we have today and in the near future.

\begin{acknowledgements} 
The authors wish to thank the entire \emph{Kepler} team, without whom these results would not be 
possible. Funding for this Discovery mission is provided by NASA's Science Mission Directorate. 
SoHO is a mission of international cooperation between ESA and NASA.This research was supported
in part by the Spanish National Research Plan under projects AYA2010-17803 and AYA2012-39346-C02-02
of the Spanish Secretary of State for R\&D\&i (MINECO). 
RAG received funding from the CNES GOLF and CoRoT grants at CEA and the ANR (Agence Nationale 
de la Recherche, France) program IDEE   ($n^\circ$ ANR-12-BS05-0008) ``Interaction Des \'Etoiles et des Exoplan\`etes''.
JB acknowledges support from CNES.
This research has also received funding from the European Communitys Seventh Framework Programme 
[FP7/2007-2013] under grant agreement No. 312844 (SPACEINN).
\end{acknowledgements} 

\bibliographystyle{aa}

\bibliography{./BIBLIO}

  \end{document}